\documentclass[12pt]{article}
\usepackage{epsfig}

\date{}

\newcommand{\beeq}{\begin{equation}}
\newcommand{\eneq}{\end{equation}}
\newcommand{\be}{\begin{eqnarray}}
\newcommand{\ee}{\end{eqnarray}}
\newcommand{\bpic}{\begin{picture}}
\newcommand{\epic}{\end{picture}}

\def\la{\raise.16ex\hbox{$\langle$} \, }
\def\ra{\, \raise.16ex\hbox{$\rangle$} }

\def\psibar{ \psi \kern-.65em\raise.6em\hbox{$-$} }
\def\mbar{ m \kern-.78em\raise.4em\hbox{$-$}\lower.4em\hbox{} }

\def\n@space{\nulldelimiterspace=0pt \mathsurround=0pt }
\def\huge#1{{\hbox{$\left#1\vbox to 20.5pt{}\right.\n@space$}}}

\def\myskip{\noalign{\kern 8pt}}
\def\myeqspace{\noalign{\kern 10pt}}

\def\boxit#1{$\vcenter{\hrule\hbox{\vrule\kern3pt
    \vbox{\kern3pt\hbox{#1}\kern3pt}\kern3pt\vrule}\hrule}$}
\def\bigbox#1{$\vcenter{\hrule\hbox{\vrule\kern5pt
     \vbox{\kern5pt\hbox{#1}\kern5pt}\kern5pt\vrule}\hrule}$}

\def\ignore#1{{}}

\begin{document}

\bibliographystyle{unsrt}
\footskip 1.0cm

\thispagestyle{empty}

\begin{flushright}
{\sc  LBNL 42696} \\
\end{flushright}

\vspace{1.0in}

\begin{center}{\Large \bf { Small $x$ Gluons in Nuclei and Hadrons }}\\

\vspace{1in}
{\large  Jamal Jalilian-Marian and Xin-Nian Wang}\\

\vspace{.2in}
{\it  Nuclear Science Division, Lawrence Berkeley National Laboratory, 
          Berkeley, CA, USA}\\

\end{center}

\vspace*{25mm}


\begin{abstract}
\baselineskip=18pt
We numerically study the effects of high gluon density at small $x$
on the evolution of gluon distribution function in both hadrons
and nuclei. Using a newly derived, Wilson renormalization group-based
evolution equation which includes $n$ to $1$ gluon ladder fusion, we 
find significant reduction in nuclear gluon distribution function for 
large nuclei at zero impact parameter at energies relevant for RHIC
and LHC experiments.  

\end{abstract}

\vspace*{5mm}

 
\newpage



\normalsize
\baselineskip=22pt plus 1pt minus 1pt
\parindent=25pt

\section{Introduction}

Gluons are the most abundant partons in hadrons and nuclei at small
$x$ (high energy) and as such, will determine the behavior of
many physical observables such as hadronic/nuclear cross sections
through their initial distribution. Once the initial distribution
of gluons in a hadron or nucleus is known, its change with $x$ and
$Q^2$ can be predicted using the powerful machinery of perturbative
QCD through the standard  QCD evolution equations such as DGLAP \cite{dglap}
and BFKL \cite{bfkl}. Both DGLAP and BFKL evolution equations can describe the
available experimental data quite well in a fairly broad range of
$x$ and $Q^2$ with appropriate parameterizations. It is interesting
to note that both equations predict a sharp growth of the gluon
distribution function as $x$ grows smaller which is clearly seen in
the Deep Inelastic Scattering (DIS) experiments at HERA (see 
\cite{cdr} for a recent review).

This sharp growth of the gluon distribution function will have to
eventually slow down in order to not violate unitarity 
(Froissart \cite{frois})
bound on physical cross sections. Gluon recombination is believed to
provide the mechanism which is responsible for this slow down 
or a possible saturation of the gluon distribution function at small
$x$. In other words, the number of gluons at small $x$ will be so large
that they will spatially overlap and therefore, gluon recombination
will be as important as gluon splitting and the standard evolution 
equations like DGLAP will have to be modified in order to take this
into effect. A first step in this direction was taken in \cite{glr}
by Gribov, Levin and Ryskin (GLR) who suggested the form of the first 
non-linear correction to DGLAP
and identified the diagrams which contribute. In \cite{mq}, Mueller
and Qiu (MQ) made a 
Gaussian like ansatz for the gluon $4$-point function
which was taken to be proportional to the square of the gluon distribution
function ($2$-point function). They then proceeded to calculate 
the numerical coefficient of the non-linear term.

It was shown in \cite{mq} that assuming reasonable values for 
$R$, a phenomenological parameter which could be taken to be either
the proton or valence quark radius, gluon recombination effects
were negligible in hadrons for not too small values of $x$ but could
be significant for large nuclei. It is perhaps helpful to mention that
GLR/MQ and gluon recombination based approaches in general are 
formulated in the infinite momentum frame. There has been much
work inspired by the approach of GLR/MQ which show that
gluon recombination leads to saturation of gluon density at small $x$
\cite{agl}. However, GLR/MQ approach includes only the first 
non-linear term in the evolution equation and will not be valid 
at very small $x$ where contribution of higher order terms will be 
as important as the first order correction and 
one will need to include them as well. Also, in the original approach 
of GLR/MQ there is no information about the impact parameter dependent 
gluon distributions and one typically assumes a factorization of the 
impact parameter and the usual gluon distribution function at all $x$.

In \cite{jkw,jklwsd}, a new evolution equation for gluon 
distribution function (or any gluonic $n$-point function) was 
derived which
is valid in the small $x$ region. It was shown in \cite{jklwbfkl}
that the new equation reduces to all the standard evolution equations
like BFKL, DLA DGLAP and GLR/MQ in the low density limit. 
In \cite{jklwdll} double leading log limit of the new equation 
was considered and a closed form was obtained which generalizes 
the GLR/MQ equation. It was shown that this new equation slows
down the growth of gluon distribution function at small $x$ 
consistent with unitarity limits. In this paper, we will numerically
solve this equation and investigate high gluon density effects on the
evolution of gluon distribution. There are many interesting 
situations where understanding these effects should be useful.

Mini-jet production at high energy is an example where high
gluon densities will play an important role. Mini-jets will 
be important at RHIC and will dominate at LHC over soft
phenomena. Nuclear shadowing of initial gluon distribution
and high gluon density could significantly reduce the initial
mini-jet and total transverse energy production. Such reduced
initial energy density will also affect the subsequent parton
thermalization.

Another example is heavy quark production where high 
gluon density effects may make a dramatic difference 
specially at LHC. Since the probability for making a heavy quark 
pair is proportional to square of gluon density, any depletion in 
number of gluons will make a significant difference in the number
of heavy quark pairs produced.

In section $2$, we discuss the relation between gluon distributions
in nuclei and hadrons followed by a brief description of  Wilson
renormalization group and effective action approach to high
density/small $x$ QCD. In section $4$, We outline the semi-classical
approach for solving the general equation and use numerical methods
to solve them. We finish by a discussion of our results and
their experimental implications as well as the limitations of our
approach.  

\section{Gluons in Hadrons and Nuclei} 

Gluon distribution function in a hadron, $xG(x,Q^2)$, has been 
studied quite extensively. Theoretically, once the initial distribution
function at a given scale $x_0$ and $Q_0^2$ is known, one can 
calculate the distribution function at a different scale  
$x$ and $Q^2$ using the standard perturbative QCD-based evolution
equations, for instance, the DGLAP equation. However, the initial 
distribution is non-perturbative
and has to be supplied as an input to the evolution equation and
is usually taken from parameterized experimental data. Alternatively,
one can use the BFKL equation to study evolution of gluon distribution
function with $x$. It is interesting to notice that both of these
evolution equations predict a sharp rise of the gluon distribution
function which would eventually lead to violation of unitarity 
(Froissart bound \cite{frois}).

Gluon distribution function of a proton can be 
measured indirectly in DIS experiments at HERA and elsewhere
by measuring virtual photon-proton cross section 
$\sigma_{\gamma^{*} p}$ where
\be
\sigma_{\gamma^{*} p} \sim {\alpha_s \over Q^2 } xG(x,Q^2).
\label{eq:F2}
\ee
It should be kept in mind that eq. (\ref{eq:F2}) is a leading twist
relation and will break down when we consider higher twist terms
in the evolution of the gluon distribution function. 
The theoretically 
predicted sharp growth of the gluon distribution function with
$x$ is observed experimentally at all $Q^2$ as well as at fixed 
(and small ) $x$ with increasing $Q^2$. Even though DGLAP evolution
equation fits the data very well, one can also use BFKL to 
explain the data and as of now, one can not experimentally
distinguish between the two scenarios.

This sharp growth of gluon distribution function is expected
to slow down eventually due to mutual interactions between
gluons when they start to spatially overlap. This is usually
referred to as saturation of gluon density and will happen
when probability of two gluons recombining into one is as large
as the probability for a gluon to split into two gluons. In other
words, when
\be
{\alpha_s \over Q^2}{xG \over \pi R^2} \sim 1
\ee
one has to include recombination effects which are neglected
in DGLAP and BFKL evolution equations.

To illustrate this, one may  write a generic evolution equation
for gluon distribution function in the high density region as
\be
{\partial^2 \over \partial y \;\partial \xi} xG \sim
\;\bigg(\sum^{\infty}_{n=0} \bigg[-{\alpha_s \over Q^2}
{xG \over \pi R^2}\bigg]^n\bigg) \; xG
\label{eq:illus}
\ee   
where $y\equiv\ln 1/x$ and $\xi\equiv \ln Q^2$. The first term of the sum 
on the right hand side of 
eq.(\ref{eq:illus}) is just the DGLAP equation whereas the second
term is referred to as GLR/MQ since it was first investigated in \cite{glr}
and its numerical coefficient was calculated by Mueller and Qiu \cite{mq}. 
An exact and formal 
evolution equation for gluon distribution function to all
orders in gluon density in the high density (small x) region
was first derived in \cite{jkw} in the infinite momentum frame. 
Whereas BFKL equation can be thought of as a ladder of reggized
gluons, the general equation derived in \cite{jkw} can be thought
of as having $n$ ladders of reggized gluon fusing into one and in this
sense, it is the generalization of BFKL equation appropriate for
the high gluon density region. 

Until recently, there was no evidence that high density
effects in a hadron were experimentally relevant even at the smallest $x$ 
(highest energies) achieved at HERA in agreement with estimates
made in MQ if one assumed reasonable values for the hadron/quark 
radius $R$. It should be 
emphasized that $R$ is a purely phenomenological parameter and
can not be derived from perturbative QCD. There is a very recent
report on slope of the proton structure function $F_2^p$ from HERA
\cite{cald} which may be an indication of importance of higher twist
effects in proton \cite{gdg}.

Gluon distribution function in a nucleus is intimately related
to that in a hadron. Typically, one assumes that nucleus is a 
weakly bound system of nucleons so that one can neglect 
inter nucleon forces which is equivalent to taking the nucleus to
be a dilute system of nucleons in the transverse plane. This is a 
good approximation for hard processes in high energy nuclear collisions
under normal conditions. 
Furthermore, if one assumes that density of gluons in a hadron is 
low, then one can simply relate distribution of gluons in nuclei
and hadrons by 
\be
xG^A(x,Q^2) = A\; xG(x,Q^2)
\ee
where $A$ is the atomic mass number, $xG^A(x,Q^2)$ and $xG(x,Q^2)$
are the nucleus and hadron (proton) gluon distribution functions.

Even if high density effects are not well established in hadrons, 
they are expected to be much more important for heavy nuclei. Non-linear 
terms in the evolution equation for gluon
distribution function in a nucleus become appreciable at a larger 
$x$ (lower energy) than for hadrons. In this sense, nucleus can be 
thought of as 
an amplifier of non-linear effects in QCD. These high density effects 
may very well be present in experiments planned at RHIC and LHC which 
underscores the crucial importance of a theoretically well-defined
approach to nuclear gluon distribution function. Also, having
nuclear beams at HERA would be of great help pinning down these effects
and would be complementary to experiments planned for
RHIC and LHC.

One of the advantages of our approach is that it can be used to
investigate the gluon distribution function in both hadrons
and nuclei and its impact parameter dependence without any
assumptions on the form of impact parameter at small $x$.
This will allow a systematic and rigorous determination of the change
in the impact parameter of nuclear gluon distribution function with energy.
To our knowledge, this is the first time that $x$ and $Q^2$ dependence 
of nuclear gluon distribution function as well as its impact parameter
dependence have been derived from QCD in the high density region. 

\section{The General Evolution Equation} 

In this section we will briefly review the Wilson renormalization
group and effective action approach to small $x$ QCD as developed
in \cite{jkw}-\cite{jklwdll}, \cite{mv}-\cite{jkmw}. To make this
paper self-contained, we will include some of the results 
reported above as needed. A few years 
ago, McLerran and Venugopalan \cite{mv} 
suggested that for very large nuclei and/or at very small $x$ one
can use weak coupling, semi-classical methods to calculate 
structure functions. They considered a large nucleus in the infinite
momentum frame and argued that as long as number of valence quarks
per unit area per unit rapidity is large, they can be treated
as static, classical sources of color charge to which the long
wavelength gluonic fluctuations (small $x$ gluons) couple. In
order to perform color averaging over the hadron/nucleus state,
they assumed a Gaussian weight for color configurations. They
proceeded to solve Yang-Mills equations of motion and calculated
the gluon distribution function in lowest order in the coupling
constant. Quantum corrections to the classical result were 
computed in \cite{ajmv} and analogous to standard perturbation
theory, large logarithmic factors ($\ln 1/x$) were encountered 
which necessitated a formalism which would resum these large 
logarithmic factors in presence of a non-trivial background 
(classical) field. 

In \cite{jklwbfkl}, McLerran-Venugopalan action was generalized 
as the following
\be
\label{action}
S &=& -{1\over 4} \int d^4 x G_a^{\mu\nu}G^a_{\mu\nu}  
+ i \int d^2 x_\perp F[\rho ^a(x_\perp)] \nonumber\\
&+& {{i}\over{N_c}} \int d^2 x_\perp dx^-
\delta (x^-)
\rho^{a}(x_\perp) {\rm tr}T_a 
W_{-\infty,\infty} [A^-](x^-,x_\perp)
\nonumber 
\ee
where $W$ is the Wilson line in the adjoint representation along the
$x^+$ axis
\be
W_{-\infty,\infty}[A^-](x^-,x_t) = P\exp \bigg[-ig \int dx^+
A^-_a(x^+,x^-,x_t)T_a \bigg]. 
\ee
The nucleus/hadron is represented by an ensemble of color charges
localized in the plane $x^-=0$ with the (integrated across $x^-$)
color charge density $\rho(x_\perp)$. Statistical 
weight of a configuration $\rho(x_\perp)$ is given by 
\be
Z=\exp \{-F[\rho]\}
\ee
In light cone gauge $A^+=0$ and at the tree level, the
chromoelectric field is determined by the color charge density through
the equations
\be
G^{+i}={1\over g}\delta(x^-)\alpha_i(x_\perp)
\label{chrom}
\ee
where the two dimensional vector potential $\alpha_i(x_\perp)$ is "pure
gauge" and is related to the color charge density by
\be
&&\partial_i\alpha^a_j-
\partial_j\alpha^b_i-f^{abc}\alpha^b_i\alpha^c_j=0\nonumber \\
&&\partial_i\alpha^a_i=-\rho^a
\label{sol}
\ee

One can then consider quantum fluctuations in background of this
classical field and separate hard and soft modes (in light cone 
longitudinal momenta) of the fluctuations, keeping terms quadratic
in hard fluctuations. Integrating out the hard modes generates the 
renormalization group equation which has the form of the evolution 
equation for the statistical weight $Z$
\cite{jkw} 
\be
{d\over dy}Z= \alpha_s \left\{{1\over 2}{\delta^2
\over\delta\rho(u)\delta\rho(v)}\left[Z\chi(u, v) \right] -{\delta
\over\delta\rho(u)}\left[Z\sigma(u)\right]\right\}
\label{final}
\ee
In the notation used in Eq.~(\ref{final}), both $u$ and $v$
stand for pairs of color index and transverse coordinate with
summation and integration over repeated occurrences implied.  The
evolution in this equation is with respect to the rapidity $y$,
related to the Bjorken $x$ by $y=\ln 1/x$.

The quantities
$\chi[\rho]$ and $\sigma[\rho]$ have the meaning of the mean
fluctuation and the average value of the extra charge density induced
by the hard modes of quantum fluctuations. They are
functionals of the external charge density $\rho$. The explicit
expressions have been given in \cite{jkw}.

Using this equation for the statistical weight $Z$, one can derive
evolution equations for n-point functions of gluon field \cite{jkw}. 
In \cite{jklwdll}, double leading log limit of the 2-point function
(gluon distribution function) was investigated and shown to be
\be
\frac{d}{d y}<\alpha_i^a(X)\alpha_i^a(Y)>=4\alpha_s\left[<X|
\frac{\alpha^2}{\partial_{\perp}^2 + 2\alpha^2}|Y>\right]^{aa}
\label{eq:twop}
\ee

In the high density limit where $\alpha^2 \gg \partial^2_{\perp}$, one can
neglect the derivative term in the denominator above and the
right hand side is a constant which leads to the gluon distribution
function growing only logarithmically with $x$ (energy) consistent
with unitarity. In the low density limit where 
$\alpha^2 \ll \partial_{\perp}^2$, one can expand the denominator in the
above equation. The first term of the expansion gives the 
DGLAP equation. Furthermore, if one assumes a factorization of the
$4$-point function in terms of the $2$-point function (as assumed
by GLR/MQ), one recovers the GLR/MQ equation \cite{mq}.  This is equivalent
to ignoring all correlations between gluon fields except that
they are constrained to be in an area of $\pi R^2$. With these
assumptions, one can actually perform the color averaging in 
(\ref{eq:twop}) which leads to
\be
\label{model}
{\partial^2 \over \partial y \partial \xi}\;xG(x,Q,b_\perp)=
\frac{N_c(N_c-1)}{2}\ Q^2\bigg[1 - 
{1 \over \kappa} \exp({1\over \kappa}) E_1({1\over \kappa})\bigg]
\label{eq:general}
\ee
where 
\be
\kappa={2 \alpha_s \over \pi (N_c -1)Q^2} xg (x,Q,b_\perp)
\ee
and ${\rm E_1}(x)$ is the exponential integral function defined 
as \cite{abr}
\be
{\rm E_1}(x)=\int_{0}^\infty\! dt\ {e^{-(1+t)x} \over 1+t},\ \ \ \ x>0
\ee 
In the low density limit, one can expand equation (\ref{eq:general}). 
Keeping the first term, we get
\be
{\partial^2 \over \partial y \partial \xi}\;xG(x,Q,b_\perp)=
{N_c\alpha_s \over \pi}\;xG(x,Q,b_\perp)
\label{eq:dglap}
\ee
which is the DLA DGLAP (small $x$ limit of DGLAP) equation. In the high
density limit, eq.  (\ref{eq:general}) gives
\be
{\partial^2 \over \partial y \partial \xi}\;xG(x,Q,b_\perp)=
{N_c (N_c -1) \over 2} Q^2
\ee
which leads to a gluon distribution of the form
\be
xG(x,Q,b_{\perp}) \sim Q^2 \ln 1/x.
\label{eq:asymp}
\ee

Let's consider the impact parameter dependent gluon distribution
function $xG(x,Q,b_{\perp})$ which is related to gluon
distribution function $xG(x,Q)$ by
\be
xG(x,Q)= \int d^2 b_{\perp} \;xG(x,Q,b_{\perp}).
\ee
It is usual to factor out the impact parameter dependence of
the distribution function and write 
$xG(x,Q,b_{\perp})=S(b_{\perp})\;xG(x,Q)$
where $S(b_{\perp})$ is the nucleus/nucleon shape function and can be
taken to be a Gaussian 
\be
S(b_{\perp})= {e^{-b_{\perp}^2/ R^2} \over \pi R^2} 
\ee
so that $\int d^2b_{\perp} \;S(b_{\perp}) =1$. This factorization 
introduces
the phenomenological parameter $R$ which is taken to be the
nuclear/hadronic radius. As long as one is using the DGLAP evolution
equation, this parameter does not come into play since DGLAP
equation is linear in gluon density. However, as we consider
the first non-linear term in the evolution equation as in 
GLR/MQ equation, parameter $R$ becomes relevant and one needs
to define it precisely. Since in GLR/MQ impact parameter dependence 
is factorized, one can only make plausible estimates of $R$. 

In general, this factorization of impact parameter will break down
with evolution in $x$ and $Q^2$ simply because gluon densities are
expected to be higher in the central ($b_{\perp}=0$) region than the 
peripheral ($b_{\perp} \sim R$) region and so therefore will evolve
differently. This would lead, in the general case, to a breakdown
of factorization of impact parameter and Gaussian ansatz for the
nucleus/nucleon shape function. In the present case where we
are working in the double logarithmic region, the Gaussian ansatz
for the shape function should still hold but the area (or radius $R$)
would change with  $x$ and $Q^2$. This basically amounts to the rise 
of perturbative
cross sections with energy. In this work, we factorize the impact 
parameter only at the starting point of our evolution  $x_0$ and 
$Q^2_0$ where non-linear effects are believed to be experimentally absent.
The evolution equation will then predict the change of this "area" with
energy. 

\section{Solving the General Equation}

In this section we will outline the procedure to numerically solve
the general equation (\ref{eq:general}). We will use the method of
characteristics which converts a partial differential equation to a 
set of coupled ordinary differential equations \cite{sned}
(see also \cite{agl,ck,eqw} for an illustration of this method). 
We will
also use MQ \cite{mq} normalization of 4-point function in terms of 2-point
functions in order to facilitate comparison of our results with those
where one includes only the first non-linear term in the evolution
equation. This amounts to a simple rescaling of our gluon distribution
function 
\be
xG(x,Q,b_{\perp}) \; \rightarrow 
{N_c(N_c-1)\pi^3 \over 6}\;xG(x,Q,b_{\perp}). 
\ee 
In the following, we will closely follow the derivation of Ayala et al.
in \cite{agl} and  rewrite (\ref{eq:general}) in terms of the
density factor $\kappa$ so that one gets
\be
{\partial^2 \over \partial y \partial \xi}\;\kappa +
{\partial \over \partial y}\;\kappa = {N_c\alpha_s \over \pi}
\bigg[1 - {1 \over \kappa} Exp({1\over \kappa}) 
E_1({1\over \kappa})\bigg]
\label{eq:keq}
\ee
where the rescaled $\kappa$ is now
\be
\kappa={N_c \alpha_s \over \pi}{\pi^3 \over 3 Q^2} xG (x,Q,b_\perp).
\ee

In the semi-classical approximation, one can write the solution to
(\ref{eq:keq}) as
\be
\kappa \equiv e^{S}
\label{eq:semic}
\ee
and neglect 
\be
{\partial^2 S\over \partial y \partial \xi}\; \ll 
{\partial S \over \partial y}\;{\partial S \over \partial \xi}.
\nonumber
\ee
Defining ${\partial S \over \partial y}\equiv \omega$ and
${\partial S \over \partial \xi}\equiv \gamma$, we get
\be
\omega (\gamma +1) = \Phi (S)
\label{eq:shorteq}
\ee
where
\be
\Phi(S)\equiv {N_c \alpha_s \over \pi}\; e^{-S} 
\bigg[1 - e^{-S + e^{-S}} E_1(e^{-S})\bigg]
\label{eq:Phi}
\ee
In terms of these variables, the set of characteristic equations 
become
\be
{dS \over dy}= {2\gamma +1 \over (\gamma +1)^2}\Phi\;\;\;\;\;\;\;\;
{d\xi \over dy}= {1\over (\gamma +1)^2}\Phi\;\;\;\;\;\;\;\;
{d\gamma \over dy}= {\gamma \over \gamma +1}
{\partial\Phi \over \partial S}.
\label{eq:char}
\ee
Notice that these equations are identical in form to those in
\cite{agl} except that our function $\Phi (S)$ is different
and will therefore result in different solutions.

In order to solve these equations, We need some initial conditions.
Since they are first order ordinary differential equations, we will
need to specify their initial values denoted by $S_0$, $\gamma_0$
and $\xi_0$ at some initial point $y_0$. In order to clarify these
initial conditions, it is helpful to write them explicitly 
in terms of the gluon distribution function
\be
S_0&=& \ln \bigg[{N_c \alpha_s \over \pi} {\pi^3 \over 2Q^2} 
\;xG(x_0,Q_0,b_{\perp})\bigg]\nonumber\\
\gamma_0&=&{\partial \over \partial \xi} \;
\ln xG(x,Q,b_{\perp})|_{x_0,Q_0} -1\nonumber\\
\xi_0&=&\ln Q^2_0
\label{eq:initial}
\ee
We will choose the initial $x_0$ and 
$Q_0$ such that the non-linear terms are negligible in the 
evolution equation. For a nucleon, this requirement is not
very restrictive since non-linear effects are small in a broad range
of $x$ and $Q^2$. In a nucleus, however, it is known experimentally 
that there is a narrow range of $x$  such that the shadowing 
ratio $S={F^A_2 \over F^N_2} \sim 1$ so that we will restrict our
initial point $x_0$ and  to lie in this region.   
From experimental data \cite{arnetal,adametal}, it appears that 
$x_0 \sim 0.05-0.07$ is a reasonable value so that for the sake of
definiteness, we will choose $x_0=0.06$ but have checked that our 
results are not
very sensitive to variation of $x_0$ in this range. We also choose
$Q_0=0.7$ (in practice, most characteristic lines start at a higher
$Q_0$) for the following reasons: quite surprisingly, all HERA data can
be explained by starting at such low values of $Q_0$ so that
perturbative QCD seems to hold at such small values in DIS (GRV 
parameterization of parton densities start at $Q_0=0.5 GeV$). 
Also, since we are using the method of characteristics to solve these 
equations, it is useful to start from a low $Q_0$ in order to
be able to find the characteristic lines in a broad range of $x$.
Most importantly, we want to get an upper limit on the amount 
of perturbative shadowing generated so that it is helpful to 
start from as low virtualities as allowed by perturbative QCD.

Having chosen our initial point $x_0$ and $Q_0$, we use CTEQ
parameterization of the proton gluon density to get $x_0 G^N(x_0,Q_0)$.
Also, at the initial point $x_0$ and $Q_0$, we assume that 
factorization of the impact parameter is valid as discussed in
some length earlier
\be
x_0G(x_0,Q_0,b_{\perp})= S(b_{\perp})x_0G(x_0,Q_0).
\nonumber
\ee

Putting everything together, at the initial point  $x_0$ and $Q_0$, 
the impact parameter dependent gluon distribution function in nuclei
and hadrons can be written as
\be
x_0 G^A(x_0,Q_0,b_{\perp}) = A {e^{-b_{\perp}^2/R_A^2} \over \pi R_A^2} 
 x_0 G^N(x_0,Q_0)
\ee
and
\be
x_0 G^N(x_0,Q_0,b_{\perp}) =  {e^{-b_{\perp}^2/R^2} \over \pi R^2} 
x_0 G^N(x_0,Q_0)
\ee
where $R^2_A$ and $R^2$ are the nuclear and hadronic areas at the
initial point $x_0$ and $Q_0$. This completely fixes our initial 
conditions for solving the set of coupled ordinary differential 
equations in (\ref{eq:char}).

We would like to emphasis that the factorization of the impact
parameter dependent distribution into a Gaussian shape function 
$S(b_{\perp})$
and the standard gluon distribution function $xG(x,Q)$ is done only
at the initial point and in principle will not hold when one goes
to small $x$ where solution of the evolution equation will determine
its functional form. Here, we will mostly work at zero impact 
parameter since that is where gluon densities and hence non-linearities 
are most important but we intend to investigate impact parameter
dependence of our results in more detail in future work.

Having determined our initial conditions, we use the $4$th order
Runge-Kutta method to solve the set of characteristic equations
in (\ref{eq:char}). Some of the characteristic lines are shown in
Figure ($1$) for illustration. All the lines start at $x_0=0.06$
and end at $Q=10 GeV$. In order to find the gluon distribution
function $ x G(x,Q,b_{\perp})$ at a given $x$, one would need
to vary the initial $Q_0$ until the corresponding characteristic
passing through the given $x$ and $Q$ is found. For the range of
$x$ and $Q$ considered here, variation of $Q_0$ is between 
$0.8 - 1.5 GeV$ so that even though the evolution formally starts
at a low value of $Q_0=0.7 GeV$, the actual initial $Q_0$ is higher. 
 
\begin{figure}[htp]
\centering
\setlength{\epsfxsize=10cm}
\centerline{\epsffile{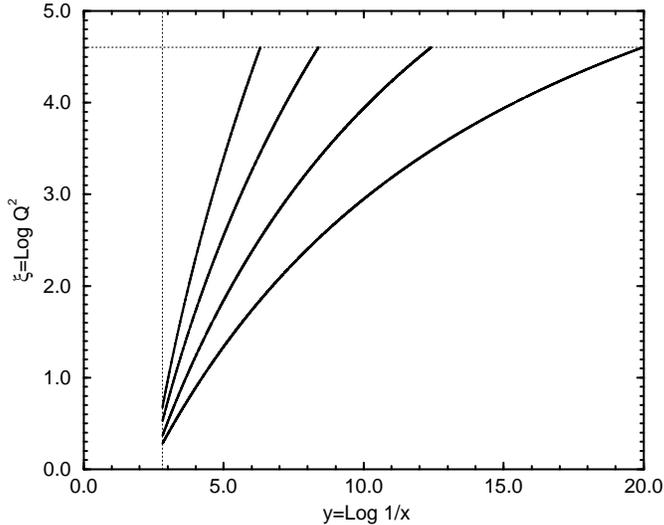}}
\caption{Some characteristic lines of equation (\ref{eq:char}) starting
at $x_0=0.06\; (y_0=2.81)$. } 
\label{fig:xsivsy}
\end{figure}

In Figures ($2$) and ($3$), we show the ratio
\be
R(x,Q,b_{\perp})={xG^{JKLW}(x,Q,b_{\perp}) \over xG^{DGLAP}(x,Q,b_{\perp})}
\label{eq:ratio}
\ee
at $b_{\perp}=0$ for both $A=1$ and $A=200$ at $Q=2 \;GeV$ and 
$Q=5 \;GeV$. We have also taken $R_A=5 \;fm$, $R=1 \;fm$ and 
$\alpha_s=.25$. Here, $xG^{JKLW}$ refers to solution of equation 
(\ref{eq:general}) while  $xG^{DGLAP}$ is the solution of
(\ref{eq:dglap}). For a proton, we get a $15-20\%$ reduction in
the number of gluons at  $x\sim 10^{-4}$  as compared to DLA DGLAP
while for a Gold or Lead nucleus, there is a $50-55\%$ reduction
at $x\sim 10^{-4}$. It is expected that these results will have
some dependence on the numerical values of the hadron or nucleus
radius as well as the coupling constant $\alpha_s$. For example, 
one can find values like $R=.5 fm$ for radius of proton and so on in 
the literature. Also, one could use a more realistic shape function 
like Woods-Saxon rather than a Gaussian but these changes are not 
expected to make more than a few percent change in our results. In
this work, we are interested in the overall features of the 
non-linear effects and will investigate these details later.

\begin{figure}[htp]
\centering
\setlength{\epsfxsize=10cm}
\centerline{\epsffile{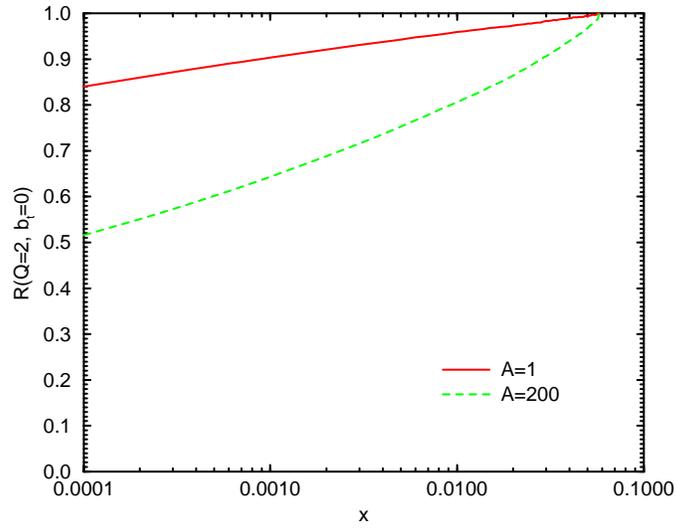}}
\caption{$R(x,Q,b_{\perp})$, as defined in (\ref{eq:ratio}) at
$Q=2 \;GeV$ and $b_{\perp}=0$.} 
\label{fig:nonlinearQ2}
\end{figure}

\begin{figure}[htp]
\centering
\setlength{\epsfxsize=10cm}
\centerline{\epsffile{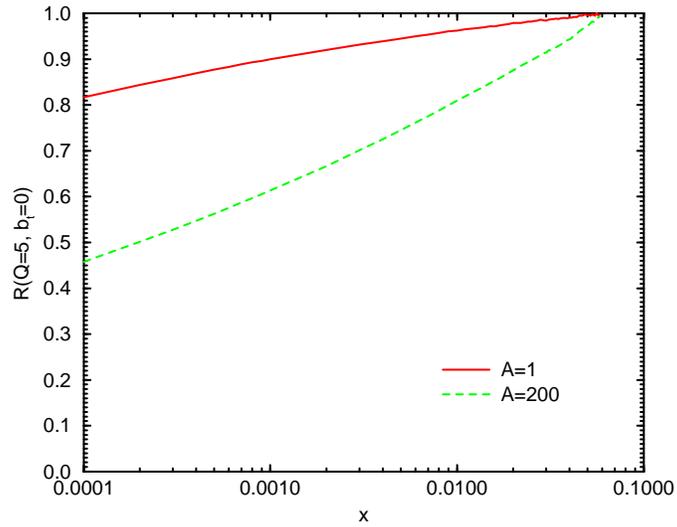}}
\caption{Same as in Figure (\ref{fig:nonlinearQ2}) at $Q=5 \;GeV$.} 
\label{fig:nonlinearQ5}
\end{figure}

In Figure $4$, we show the $Q$ dependence of
this ratio at different $x$ for $A=200$. To make comparison easier,
we normalize $R=1$ at $Q=1$. For a proton, the $Q$ dependence was 
found to be negligible and is not shown.

\begin{figure}[htp]
\centering
\setlength{\epsfxsize=10cm}
\centerline{\epsffile{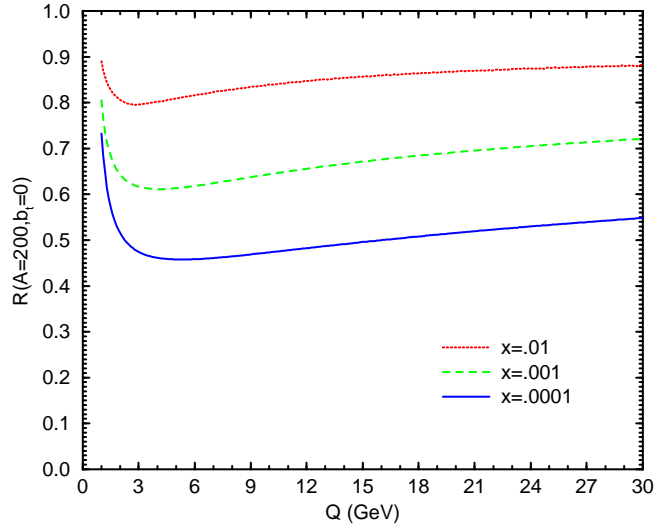}}
\caption{$R(x,Q,b_{\perp})$  as a function of $Q$ at 
different $x$ for $A=200$.} 
\label{fig:nonlinearA200x}
\end{figure}

As is seen, the non-linearities become less important at higher
$Q$. However, the low $Q$ behavior is peculiar since one expects
a monotonous decrease of $R$ with decreasing $Q$ while it is seen
to eventually increase with decreasing $Q$ and tend to $1$. This
dip in $Q$ is a consequence of using the method of characteristics 
to solve partial differential equations and is not an artifact of 
our formalism (see, for example, Figs. 23, 24 in \cite{agl}). 

To find the gluon distribution
function $xG(x,Q)$ at $Q$, one needs to find the characteristic line
of the corresponding evolution equation passing through that 
value of $Q$ by finding the value of $Q_0$ from which the 
desired characteristic line starts. $xG^{JKLW}$ and $xG^{DGLAP}$ 
satisfy different evolution equations and therefore have different
characteristic lines which start at different values of $Q_0$ in order
to reach the point $Q$. In other words, the initial starting $Q_0$
is never exactly the same for the two evolution equations. This would
not be important if the distribution functions at the initial
$Q_0$  were slowly varying which is not the case for $xG^{JKLW}(x,Q)$ 
since it includes high density effects. To make a completely 
self-consistent treatment of perturbative shadowing possible,
one would need to start from parameterizations of parton distributions
which include initial non-perturbative shadowing such as \cite{eskola}
rather than CETEQ, GRV or MRS. We intend to do this when we study
perturbative gluon shadowing in nuclei in the near future.

In Figures ($5$) and ($6$), we show the $A$ and $b_{\perp}$ dependence 
of our results at fixed
$Q$ for different values of $x$. As is seen, non-linear
effects set in rather quickly, specially at low $x$, after which they
increase slowly.

\begin{figure}[htp]
\centering
\setlength{\epsfxsize=10cm}
\centerline{\epsffile{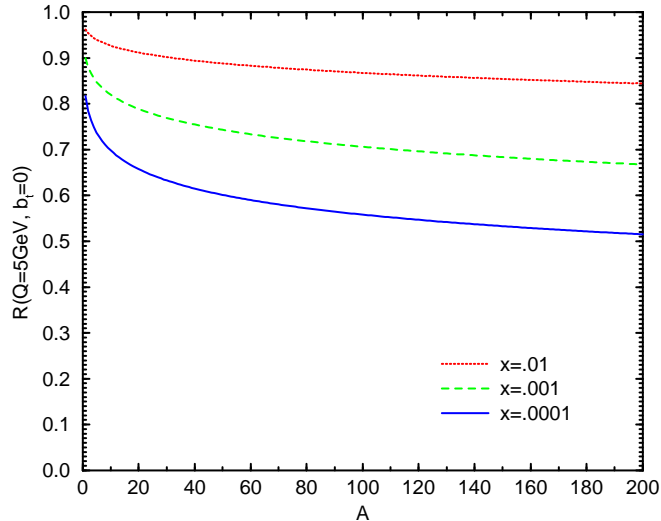}}
\caption{$R(x,Q,b_{\perp})$  as a function of $A$ at $Q=5 GeV$ for 
different $x$.} 
\label{fig:nonlinratiovsA}
\end{figure}

\begin{figure}[htp]
\centering
\setlength{\epsfxsize=10cm}
\centerline{\epsffile{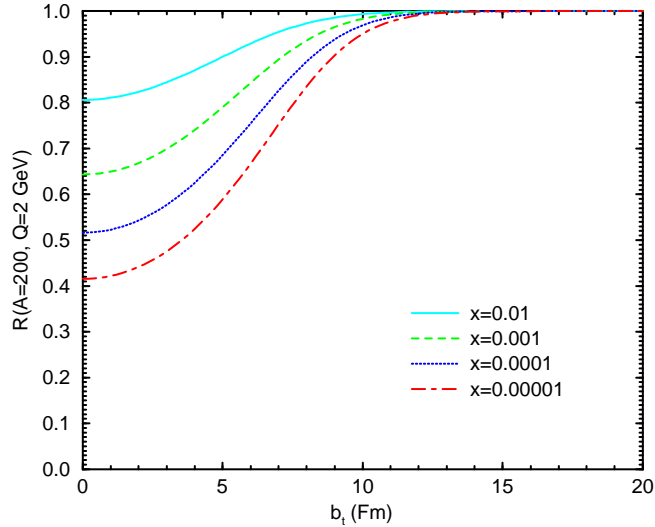}}
\caption{$R(x,Q,b_{\perp})$  as a function of $b_{\perp}$ at 
different values of $x$.} 
\label{fig:RvsbtQ2atx}
\end{figure}

These figures clearly show the importance of the
non-linear terms, specially for a large nucleus, at values of $x$
which will be reached in the upcoming experiments at RHIC and LHC.
These non-linear effects will be manifest in terms of shadowing
of nuclear gluon distribution function and will have to be taken
into account at future high energy heavy ion experiments. We are
currently investigating shadowing of nuclear gluon 
distributions \cite{jw} using this formalism.

\section{Discussion}

We have investigated the $x$ and $Q$ dependence of gluon recombination 
effects on the evolution of impact parameter dependent gluon 
distribution function in hadrons and nuclei using a general evolution
equation which takes $n \rightarrow 1$ gluon ladder fusion into
account. Using the method of characteristics, we numerically solved
the general evolution equation and found that gluon recombination effects
are very important specially for central ultra-relativistic heavy 
ion collisions coming up at RHIC and LHC.

A detailed knowledge of nuclear gluon distribution will be essential
in understanding the outcome of RHIC and LHC experiments. With our
formalism and with the solution to the general evolution equation at hand,
we can investigate several aspects of these distributions. The first 
thing to consider is shadowing of nuclear gluon distributions
using the present formalism. In addition to the usual shadowing
ratio defined as ${xG^A(x,Q) \over AxG(x,Q)}$, we can investigate
impact parameter dependence of this ratio. This will give information 
about shadowing of gluon distributions in the peripheral as well as the
central region. 
Studying $x$ and $Q$ 
dependence of the average impact parameter will determine how 
the effective area (cross section) of a nucleus changes with 
increasing energy of nuclear collisions. The A dependence of shadowing
ratio at small $x$ can also be determined rigorously without any 
model dependent assumptions \cite{mq}.
One should also integrate over the impact parameter eventually in
order to make possible a comparison with other approaches which take only
the first non-linear term into account. 
Work in this direction is in progress \cite{jw}. 

One could also consider the effect of shadowing of nuclear gluon 
distributions on physical cross sections like Drell-Yan \cite{bhq} 
and heavy quark production. To do this, one would need 
to include the relevant higher twist effects not only in the 
evolution of gluon distribution (as is done here) 
but also in the relation between the cross section (or $F_2$) and gluon
distribution function. In \cite{rl}, the all twist structure function,
$F_2$, was computed in the infinite momentum frame 
(see also \cite{agl,yk}
for a similar calculation in the lab frame) so that all one has to do is
to merge the two results \cite{jar}.
  
There are a few issues which need to be analyzed further. First, one
should investigate our choice of the initial virtuality $Q_0$ in
solving the evolution equation. We took a low value of  $Q_0=0.7 \;GeV$
(see the comments before eq. (\ref{eq:ratio}))
because it seems to be consistent with DIS data from HERA and in order
to get maximum perturbative shadowing of gluon distribution possible.
However, from Figure ($4$), we see that $R$ decreases rather 
sharply at low values of $Q$ before it starts to increase with 
increasing $Q$ as it must as a high twist effect. This means that
our treatment, strictly speaking, is not self-consistent at low values of
$Q$. In other words, starting from high values of $Q$ and decreasing
$Q$, we should have a monotonous decrease of $R$ as is seen in 
Figure ($4$). Further decrease of $Q$ leads to $R$ increasing in order
to match our initial condition that $R(x_0,Q_0)=1$. This indicates that
one should include some initial non-perturbative shadowing
so that $R(x_0,Q_0)\neq 1$. Unfortunately, this requires detailed 
knowledge of $x$, $Q$, $A$ and $b_{\perp}$ dependence of gluon
distribution function at the initial point which necessitates use of
many model dependent assumptions. There are a number of approaches 
including vector meson dominance, Pomeron exchange
models, etc. ( see \cite{sk} and references therein) which one could 
adopt to address this issue. 
This is beyond the scope of present work and will be pursued in
future. 

We are also going to study the dependence of our results
on our choice of CTEQ parameterization of gluon densities by 
repeating this calculation using other available parameterizations
like MRS and GRV. However, since we used the parameterized gluon
distributions at a fairly high initial $x_0 \sim 0.06$ and because  
parameterization dependence of gluon distribution function becomes
noticeable only at small values of $x$, we do not believe our results
will be sensitive to choice of parameterization.

Another point to keep in mind is that we have been working in the leading
log approximation and therefore have taken $\alpha_s$ to be a constant.
It is known that DLA DGLAP with fixed  $\alpha_s$ overestimates the
gluon number density at small $x$ and to get good agreement with 
experimental data from HERA, one needs to include next to leading order 
corrections to DGLAP. However, our general evolution equation is
derived in the leading log approximation and will be modified if
one goes beyond the leading log approximation. One such modification
due to next to leading order corrections is to cause running of
$\alpha_s$, but this may not be the only effect or even the dominant
effect so that to be theoretically consistent, we have kept everything 
at the leading log approximation level. However, so long as we are 
working with ratios of distributions, we think next to leading order 
corrections to this ratio will not be large and a leading order 
calculation such as this should be adequate.

Our emphasis in this work has been on theoretical self-consistency
so that we have not made any attempt to fit or reproduce experimental
data. The main reason is that until now, there has been little
effort made to calculate nuclear gluon distribution function directly from
QCD without resorting to elaborate modeling. Also, there is not much
experimental data available on nuclear gluon distribution function.
What is experimentally measured is the structure function $F_2$ and
to get the gluon distribution function, one takes logarithmic 
derivative of $F_2$
\be
xG(x,Q) \sim {d \over d\log Q^2} F_2.
\label{eq:twist}
\ee

As mentioned earlier, this is a leading twist relation and will
not be valid in a general all twist calculation such as ours.
By using eq. (\ref{eq:twist}) to extract the gluon distribution 
function experimentally, one is implicitly assuming that higher 
twist effects are
not present. To be theoretically consistent, one should figure out how
to extract gluon densities from experimental data allowing for 
higher twist effects. Until this is done, in our opinion, one should 
develop a self-consistent approach
derived from fundamental theory with well defined approximations
and with as less model dependence as possible. This is the approach 
adopted in this work.

\leftline{\bf Acknowledgments} 

We would like to thank S. Brodsky, S. Jeon, V. Koch, Y. Kovchegov, A. Kovner, 
L. McLerran, B. Mueller, Y. Pang, R. Venugopalan and R. Vogt for 
various discussions on topics related to this work. This work was
supported by the Director, Office of Energy Research, Office of High Energy
and Nuclear Physics Division of the Department of Energy, under 
contract No. DE-AC03-76SF00098 and DE-FG02-87ER40328.

\leftline{\bf References}

\renewenvironment{thebibliography}[1]
        {\begin{list}{[$\,$\arabic{enumi}$\,$]}  
        {\usecounter{enumi}\setlength{\parsep}{0pt}
         \setlength{\itemsep}{0pt}  \renewcommand{\baselinestretch}{1.2}
         \settowidth
        {\labelwidth}{#1 ~ ~}\sloppy}}{\end{list}}


\begin{thebibliography}{99}

\small

\bibitem{dglap}
V.N. Gribov and L.N. Lipatov, {\it Sov. J. Nucl. Phys. }{\bf 15}, 78
(1972); G. Altarelli and G. Parisi, {\it Nucl. Phys. }{\bf B126}, 298
(1997); Yu. L. Dokshitzer, {\it Sov. Phys. JETP }{\bf 73}, 1216 
(1997).

\bibitem{bfkl}
E.A. Kuraev, L.N. Lipatov and V.S. Fadin, {\it Sov. Phys. JETP }
{\bf 45}, 199 (1977); Ya. Ya. Balitskii and L.N. Lipatov, 
{\it Sov. J. Nucl. Phys. }{\bf 28}, 22 (1978).

\bibitem{cdr}
A.M. Cooper-Sarkar, R.C.E. Devenish and A. De Roeck, {\it Int. J. Mod.
Phys. }{\bf A13}, 3385 (1998).

\bibitem{frois}
M. Froissart, {\it Phys. Rev. }{\bf 123}, 1053 (1961).

\bibitem{glr}
L.V. Gribov, E.M.~Levin and M.G.~Ryskin, {\it Phys.\ Rep. }{\bf 100}, 1 
(1983).

\bibitem{mq} 
A.H. Mueller and J.W. Qiu, {\it Nucl. Phys. }{\bf B268}, 427 (1986);
J.W. Qiu,  {\it Nucl. Phys. }{\bf B291}, 746 (1987).

\bibitem{agl}  
A.L. Ayala, M. B. Gay Ducati and E. M. Levin,
{\it  Nucl. Phys. }{\bf B493}, 305 (1997); {\bf B511}, 355 (1998).

\bibitem{jkw} 
J. Jalilian-Marian, A. Kovner and H. Weigert,
{\it Phys. Rev. }{\bf D59}, 014015 (1999).

\bibitem{jklwsd} 
J. Jalilian-Marian, A. Kovner, A. Leonidov and H. Weigert, 
{\it Phys. Rev. }{\bf D59}, 014014 (1999).

\bibitem{jklwbfkl} 
J. Jalilian-Marian, A. Kovner, A. Leonidov and H. Weigert, 
{\it Nucl. Phys. }{\bf B504}, 415 (1997).

\bibitem{jklwdll} 
J. Jalilian-Marian, A. Kovner, A. Leonidov and H. Weigert, 
{\it Phys. Rev. }{\bf D59}, 034007 (1999).

\bibitem{cald}
A. caldwell, DESY Theory Workshop, DESY, October 1997.

\bibitem{gdg}
M.B. Gay Ducati and V. Goncalves, hep-ph/9812459.

\bibitem{mv} 
L. McLerran and R. Venugopalan, {\it Phys. Rev. }{\bf D49}, 335 (1994);
{\bf D49}, 2233 (1994).

\bibitem{ajmv}
A. Ayala, J. Jalilian-Marian, L. McLerran and R. Venugopalan,
{\it Phys. Rev. }{\bf D52}, 2935 (1995); {\bf D53}, 458 (1996). 

\bibitem{jkmw} 
J. Jalilian-Marian, A. Kovner, L. McLerran and H. Weigert,
{\it Phys. Rev. }{\bf D55}, 5414 (1997).

\bibitem{abr}
M. Abramowitz and I. Stegun, "Handbook of Mathematical Functions",
1972.

\bibitem{sned}
I.N. Sneddon, "Elements of Partial Differential Equations", 
McGraw-Hill, New York, 1957.

\bibitem{ck}
J. Collins and J. Kwiecinski, {\it Nucl. Phys. }{\bf B335}, 89 (1990). 

\bibitem{eqw}
K. Eskola, J. Qiu and X.N. Wang, {\it Phys. Rev. Lett. }{\bf 72}, 36 (1994).

\bibitem{arnetal}
M. Arneodo,  {\it Phys. Rep. }{\bf 240}, 301 (1994).

\bibitem{adametal}
M.R. Adams, et.al.,  {\it Phys. Rev. Lett. }{\bf 68}, 3266 (1992).

\bibitem{eskola}
K. Eskola, et al.,  {\it Nucl. Phys. }{\bf B535} 351 (1998). 

\bibitem{jw}
J. Jalilian-Marian and X.N. Wang, Mauscript in preparation.

\bibitem{bhq}
S.J. Brodsky, A. Hebecker and E. Quack, 
{\it Phys. Rev. }{\bf D55}, 2584 (1997). 

\bibitem{rl}
L. McLerran and R. Venugopalan, hep-ph/9809427.

\bibitem{yk}
Yu. Kovchegov, hep-ph/9901281.

\bibitem{jar}
A. Kovner and R. Venugopalan, private communications.

\bibitem{sk}
S. Kumano, {\it Phys. Rev. }{\bf C48}, 2016 (1993). 

\end{thebibliography}
\end{document}